\documentclass{aa}
\usepackage{graphicx}
\begin{document}
\title{The IMF of open star clusters with Tycho--2\thanks{Based on
    observations of the ESA Hipparcos satellite}}

\author{J\"org Sanner \and Michael Geffert}

\offprints{J\"org Sanner, \email{jsanner@astro.uni-bonn.de}}

\institute{Sternwarte der Universit\"at Bonn, Auf dem H\"ugel 71, 
           53121 Bonn, Germany}

\date{Received 14 September 2000 / Accepted 7 February 2001}

\abstract{We studied the fields of nine nearby open star clusters based on
  the Tycho--2 catalogue. We determined membership probabilities for the stars
  in the cluster fields from the stellar proper motions and used the Tycho--2
  photometry to compute the initial mass function (IMF) of the clusters from
  the main sequence turn-off point down to approx. $1 M_\odot$. We found IMF
  slopes ranging from $\Gamma=-0.69$ down to $\Gamma=-2.27$ (when the Salpeter
  (\cite{salpeter}) value would be $\Gamma=-1.35$). We also studied the
  membership of individual stars of special astrophysical interest. In some
  cases previous results had to be revised. As a by-product, we
  investigated some general properties of the Tycho--2 catalogue; we
  confirmed that the Tycho--2 proper motions show only marginal deviations
  from the Hipparcos catalogue. On the other hand, in some regions the
  completeness of the catalogue seems to decrease at magnitudes brighter than
  claimed by H{\o}g et~al.\ (\cite{tycho2}).
\keywords{open clusters and associations: general -- catalogs -- astrometry --
  stars: kinematics -- Hertzsprung-Russell (HR) and C-M diagrams -- stars:
  luminosity function, mass function}}

\authorrunning{J. Sanner \& M. Geffert}
\titlerunning{The IMF of open star clusters with Tycho--2}

\maketitle

\section{Introduction} \label{tychointro}

The Tycho--2 catalogue (H{\o}g et~al.\ \cite{tycho2}) provides positions and
stellar proper motions in conjunction with two colour photometric data. These
were computed from the original Tycho (star mapper) data obtained during the
Hipparcos mission (ESA \cite{hipp}) and ground based first epoch data,
including the Astrographic Catalogue (Wood \cite{wood}). This work
was triggered by the findings of Makarov et~al.\ (\cite{makarov}) who recently
found that the precision of the Tycho--2 proper motions are of the same order
as the Hipparcos data. With its significantly larger number of stars, the
Tycho--2 catalogue therefore is a suitable source of data for studies of
stellar kinematics. In earlier investigations Hipparcos data were used for a
search for open clusters (Platais et~al.\ \cite{platais}) and checks of open
cluster candidates (Baumgardt \cite{baumgardt}).

In this investigation we used fields of nine nearby open star clusters. We
separated the field stars from the cluster members with the help of the proper
motions provided by Tycho--2 and determined colour magnitude diagrams (CMDs)
from its photometry. From the CMDs we derived the age, distance modulus, and
reddening by fitting isochrones and compared the results with, e.g., distances
derived from Hipparcos parallaxes (van~Leeuwen \cite{vanleeuwenhipp},
hereafter referred to as vL99a). In a final step, we computed the initial mass
function (IMF) of the clusters with a maximum likelihood method. The results
of this part of the study can give us clues about the completeness of the
Tycho--2 data.

Especially for open clusters, where the contamination of the data with field
stars poses a major challenge, knowledge of membership is essential to derive
the IMF. The most precise method to determine membership is by studies of
stellar kinematics, as one can assume that all cluster stars move in the same
way, whereas the distribution of the motions for the field stars is much
wider and differently centred. From this point of view, a catalogue like
Tycho--2, including photometry and proper motions, provides an ideal
opportunity for an IMF study.

The shape of the IMF is an important parameter to understand the fragmentation
of molecular clouds and therefore the formation and development of stellar
systems. Besides studies of the Solar neighbourhood (Salpeter \cite{salpeter},
Tsujimoto et~al.\ \cite{tsuji}), work on star clusters plays an important role
(Scalo \cite{scalo1,scalo2}), as age, metallicity, and distance of all cluster
stars can be assumed to be the same.

Restricted to certain mass intervals, most studies agree that the IMF can
be described by a power law in the form
\begin{equation}
\mbox{d} \log N(m) \sim m^\Gamma \mbox{d} \log m.
\end{equation}
In this notation, the ``classical'' value found by Salpeter (\cite{salpeter})
for the Solar neighbourhood would be $\Gamma=-1.35$. We will not be able to
reach smaller masses than about $1 M_\odot$ in this study, so that,
according to the review of Scalo (\cite{scalo2}), our values for $\Gamma$ are
expected to be within the interval $[1.2;2.2]$, a range which is
valid for intermediate mass stars of $1 M_\odot \la m \la 10 M_\odot$. In
particular, the IMF can be expected to show a behaviour which is in accordance
with a single power law in this mass range.

Up to now, thorough IMF studies are available only for a few of the
nearby star clusters, e.g the Pleiades (van~Leeuwen \cite{vLpleiades},
Meusinger et~al.\ \cite{meusinger}, Hambly et~al.\ \cite{pleiadesimf}) or
Praesepe (Williams et~al.\ \cite{praesepeimf}, Pinfield et~al.\
\cite{pinf}). Some of the clusters were also studied by Tarrab
(\cite{tarrab}), but her work was based on an inhomogeneous data sample with
sometimes weak information on membership. Discrepancies of her results with
later work have been found (see, e.g., Sanner et~al.\ \cite{capaper} for
NGC\,1960 or Credner \cite{credner} for NGC\,2323), so that revisiting the
clusters studied by Tarrab (\cite{tarrab}) is promising.

Nowadays, although CCD cameras provide a good tool for IMF studies down to low
masses ($m \ll M_\odot$), the limited field size of typical imagers of the
order of arc minutes makes comprehensive studies a difficult task: Nearby
clusters have a large angular diameter in the sky, so that mosaics of lots of
different fields would have to be observed to obtain a general overview of the
stellar content of a cluster. For the determination of a cluster IMF a
snapshot of one or only few fields is not enough, since spatial
inhomogeneities (mass segregation; see, e.g., Raboud \& Mermilliod
\cite{raboud2}) are likely to falsify the results. With a uniform photometric
and astrometric all sky survey, this goal can be achieved at least down to a
reasonable mass limit.

Sect.~\ref{tychodata} contains a description of the data and methods used for
this work as well as some general results, followed in Sect.~\ref{tychoindiv}
by the results sorted by cluster. The final section contains a summary and
conclusions.

\section{Data, data analysis and general results} \label{tychodata}

Since the Tycho--2 data are claimed to be complete down to $11 \mbox{~mag}$ (see H{\o}g et~al.\ \cite{tycho2} and the discussion below), we were
restricted to a small magnitude (and hence mass) interval. To reach stars with
masses as low as possible we chose
to study nearby ($(m-M)_0 \la 7.5 \mbox{~mag}$ or $d \la 315 \mbox{~pc}$)
open star clusters only.

Robichon et~al.\ (\cite{robichon}) published a list of astrometric data of
nearby open clusters based on the Hipparcos catalogue. From this information
the nine targets listed in Table \ref{tychohaufen} were chosen. A main
criterion for the selection was a clear separation between the field and
cluster proper motion centres to make the membership determination an easy
task and to avoid the presence of numerous field stars with the same proper
motion as the cluster. A maximum of 10 to 15 stars can be expected to belong
to the field but to have (within the errors) the same proper motion as the
corresponding cluster. The Hipparcos parallaxes of seven of them were recently
studied by vL99a to derive the distances of the clusters. Similar computations
can also be found in Robichon et~al.\ (\cite{robichon}). Due to the very high
quality of the Hipparcos data and the reliability of the results, we preferred
these values for the distance moduli and reddenings of these clusters compared
with those of previous studies. The results of vL99a and Robichon et~al.\
(\cite{robichon}) agree well within the errors for our clusters, except for
Blanco\,1 with a deviation of $\Delta (m-M)_0 \approx 0.3 \mbox{~mag}$. We
will discuss this point later (Sect.~\ref{blanco1}).

\begin{table}
\caption[]{\label{tychohaufen} List of the open star clusters studied in this
  work together with their coordinates. For data on the proper motions, see
  Table \ref{tychoebs}}
\begin{tabular}{lrr}
\hline
cluster & \multicolumn{1}{c}{$\alpha_{2000}$} &
 \multicolumn{1}{c}{$\delta_{2000}$}\\
 & \multicolumn{1}{c}{[h m]} & \multicolumn{1}{c}{[$^\circ$ $\arcmin$]}\\
\hline
Blanco\,1    & $00$ $04.3$ & $-29$ $56$ \\
Stock\,2     & $02$ $15.0$ & $+59$ $16$ \\
$\alpha$ Per & $03$ $22.0$ & $+48$ $37$ \\
Pleiades     & $03$ $47.0$ & $+24$ $07$ \\
NGC\,2451\,A & $07$ $45.4$ & $-37$ $58$ \\
IC\,2391     & $08$ $40.2$ & $-53$ $04$ \\
Praesepe     & $08$ $40.4$ & $+19$ $40$ \\
IC\,2602     & $10$ $43.2$ & $-64$ $24$ \\
NGC\,7092    & $21$ $32.2$ & $+48$ $26$ \\
\hline
\end{tabular}
\end{table}

\begin{table*}
\caption[]{\label{tychoebs} List of the open star clusters studied. We give
  the centres of the proper motions of the field and cluster stars. Since the
  field proper motions were fitted by a two dimensional Gaussian function, we
  give the semi major and minor axes $a$ and $b$ and the rotation angle $\phi$
  of the $1 \sigma$ ellipse. Furthermore, we add the ratio $r$ of members to
  all stars in the studied fields. For a comparison of our Tycho--2 proper
  motion centres with the Hipparcos-based ones from Robichon et~al.\
  (\cite{robichon}), see Sect.~\ref{tychodata}}
\begin{tabular}{l|rrr|rrrrr|r}
\hline
 & \multicolumn{3}{c|}{cluster} & \multicolumn{5}{c|}{field} & \\
cluster & $\mu_\alpha \cos \delta$ & $\mu_\delta$ & \multicolumn{1}{c|}{$\sigma$} & \multicolumn{1}{c}{$\mu_\alpha \cos \delta$} & \multicolumn{1}{c}{$\mu_\delta$} & \multicolumn{1}{c}{$a$} & \multicolumn{1}{c}{$b$} & \multicolumn{1}{c|}{$\phi$} & \multicolumn{1}{c}{$r$}\\
 & \multicolumn{1}{c}{[mas yr$^{-1}$]} & \multicolumn{1}{c}{[mas yr$^{-1}$]} &
 \multicolumn{1}{c|}{[mas yr$^{-1}$]} & \multicolumn{1}{c}{[mas yr$^{-1}$]} &
 \multicolumn{1}{c}{[mas yr$^{-1}$]} & \multicolumn{1}{c}{[mas yr$^{-1}$]} &
 \multicolumn{1}{c}{[mas yr$^{-1}$]} & \multicolumn{1}{c|}{[$\deg$]} & \multicolumn{1}{c}{[$\%$]}\\
\hline
Blanco\,1    & $+19.27$ & $+2.81$ & $2.02$ & $+10.98$ & $-7.71$ & $16.48$ & $11.93$ & $+24.5$ & $11.1$ \\
Stock\,2     & $+17.15$ & $-12.94$ & $4.61$ & $+0.36$ & $-1.58$ & $6.76$ & $6.40$ & $-31.8$ & $14.0$ \\
$\alpha$ Per & $+22.63$ & $-25.74$ & $1.52$ & $-0.74$ & $-3.53$ & $13.11$ & $8.14$ & $-37.0$ & $3.7$ \\
Pleiades     & $+19.96$ & $-45.09$ & $1.78$ & $+5.43$ & $-11.42$ & $17.62$ & $14.31$ & $-53.1$ & $7.6$ \\
NGC\,2451\,A & $-21.96$ & $+14.59$ & $1.17$ & $-5.40$ & $+4.59$ & $11.21$ & $7.33$ & $-57.8$ & $5.4$ \\
IC\,2391     & $-24.64$ & $+23.25$ & $1.13$ & $-6.82$ & $+6.29$ & $12.34$ & $7.28$ & $-48.3$ & $6.0$ \\
Praesepe     & $-35.88$ & $-13.18$ & $1.93$ & $-6.40$ & $-9.20$ & $11.47$ & $11.32$ & $+6.5$ & $10.6$ \\
IC\,2602     & $-17.63$ & $+10.30$ & $1.50$ & $-8.55$ & $+2.43$ & $9.62$ & $5.68$ & $-26.6$ & $4.5$ \\
NGC\,7092    & $-8.34$ & $-19.16$ & $1.19$ & $-2.00$ & $-2.26$ & $10.93$ & $6.52$ & $+41.1$ & $2.7$ \\
\hline
\end{tabular}
\end{table*}

The Tycho project made use of its own photometric system, referred to as $B_T$
and $V_T$. These two filters are close to Johnson $B$ and $V$, respectively,
and ESA (\cite{hipp}) provides transformation equations to the Johnson system:
\begin{eqnarray}
V&=&V_T-0.090 \cdot (B-V)_T\\
B-V&=& 0.850 \cdot (B-V)_T. \label{bvtgleichung}
\end{eqnarray}
For our purposes, it was not necessary to transform the magnitudes to
$B$ and $V$: The $V$ magnitudes are influenced by errors of up to $\Delta
V=0.09 \mbox{~mag}$ -- a value which is valid for the redmost stars
(i.e. $(B-V)_T \approx 1 \mbox{~mag}$). The colours can be influenced by up
to $\Delta(B-V)=0.15 \mbox{~mag}$, however, this figure again is valid for
very red stars only. Since the colours are only required for the isochrone
fitting, which is performed on the base of the bright main sequence stars
with colours of around $(B-V)_T=0$, the deviations from the Tycho to the
Johnson system can be disregarded. We will see later that the Tycho--2
$(B_T,V_T)$ data lead to lower reddenings than several previous studies of the
same objects. This cannot be explained by not transforming the magnitudes to
the Johnson system: In addition to the arguments mentioned before, the Johnson
colours for stars with $(B-V)_T \ge 0$ are {\it bluer} than the corresponding
Tycho values according to Eq. (\ref{bvtgleichung}). This is in contrast with
the deviating values, which would all lead to a higher reddening.

After extracting sufficiently large areas around the centres of the targets
-- taking at least twice the diameter given in the Lyng{\aa} (\cite{lynga})
catalogue to avoid a biasing due to mass segregation -- we derived the
membership probabilities $P$ of the individual stars from their proper
motions. We fitted a combination of a sharp (for the members) and a wider (for
the field stars) two-dimensional Gaussian distribution to the vector point
plot diagram. The optimal parameters (location of the field and cluster proper
motion centres in the diagram and the standard deviations of the two
Gaussians) were determined with a maximum likelihood procedure as described
in Sanders (\cite{sanders}).

Due to the large fields, the vector point plot diagrams were dominated by the
field stars. For some of the clusters, this caused the software to
misinterpret the field stars as the cluster members and vice versa. Since the
stars determined as members are located within a circle or an ellipse, and the
non-members consist of all the rest, the correct membership probabilities
could not be computed as the complement of the determined ones: Any star at a
sufficient distance from the field centre would obtain a high probability with
this method, even if that star is located at the opposite end of the
vector point plot diagram, compared with the actual cluster. This problem
could be solved by removing all stars with a very large deviation from the
cluster proper motion ($>20 \mbox{~mas~yr}^{-1}$). Since the clusters were
easily detectable in the vector point plot diagrams and the Robichon et~al.\
(\cite{robichon}) data gave additional clues about their location, it can be
excluded that the region covering the target was accidentally eliminated. Also,
cross checks of the positions of the stars selected as members in the vector
point plot diagram showed that the membership determination was successful.

The vector point plot diagrams (see the diagram of the $\alpha$ Per cluster
derived from the Tycho--2 data in Fig.~\ref{alphapervppd} as an example)
showed a standard deviation of the cluster stars in the region of $1.5-2
\mbox{~mas~yr}^{-1}$. Since these values correspond to proper motion errors
rather than to internal motions, a good coincidence between our results and
the values given in H{\o}g et~al.\ (\cite{tycho2}, their Table 2) for the
Tycho--2 accuracy was found. The only exception is the cluster Stock\,2 with a
width of the distribution of approx. $4 \mbox{~mas~yr}^{-1}$. We will return
to this outlier in Sect.~\ref{stock2}.

\begin{figure*}
\includegraphics[width=\textwidth]{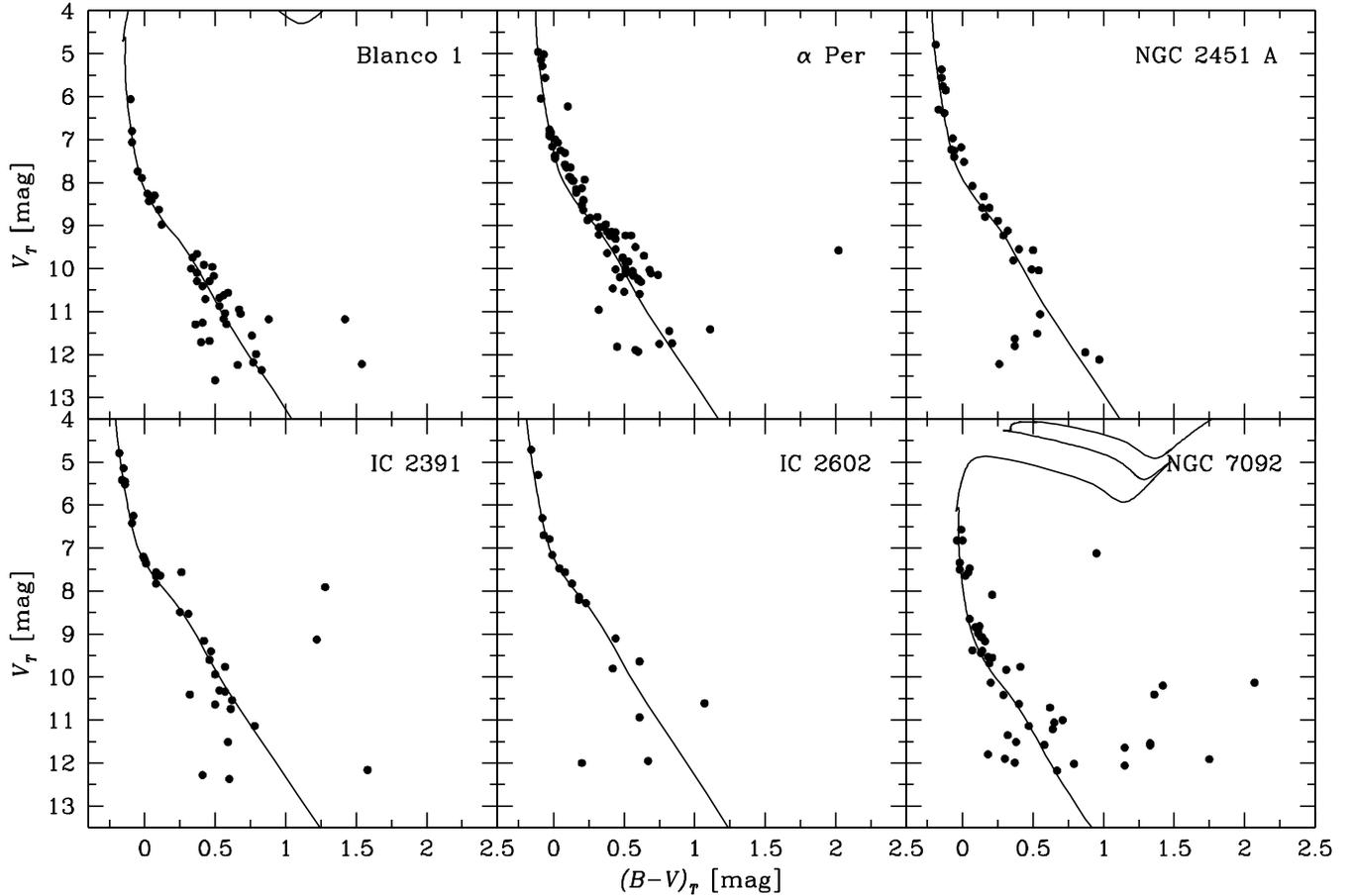}
\hfill
\caption[]{\label{tychocmds} CMDs of the members of six of the eight open star
  clusters from Tycho--2 data. We overplotted the best-fitting isochrones. The
  CMD of Stock\,2 is discussed in more detail in Sect.~\ref{stock2}, the CMD
  of the Pleiades is plotted in Fig.~\ref{pleiadescmd}. Note the lack of stars
  fainter than $V_T=8 \mbox{~mag}$ in the CMD of IC\,2602}
\end{figure*}

From the stars with a sufficiently high membership probability ($P>0.5$ to
$P>0.8$, depending on the object) we derived CMDs of the clusters. We fitted
isochrones based on the stellar models of Bono et~al.\ (\cite{isoteramo}) and
provided by Cassisi (private communication) to the CMDs to derive estimations
for the distance moduli, reddenings, and ages of the clusters. Our initial
values were those of vL99a and Robichon et~al.\ (\cite{robichon}) which in
most cases led to a sufficiently good fit. For some clusters, however, the
parameters had to be slightly adjusted, since the isochrones based on these
results clearly did not represent the CMDs well. In those cases, which will
individually be discussed in the corresponding sections, we used our own
results. The metallicity was in all cases assumed to be Solar, which is,
according to the Lyng{\aa} (\cite{lynga}) catalogue, appropriate for all
clusters. The only exception may be NGC\,2451\,A which is
metal-deficient. Values for its metallicity as given in the literature range
from $Z=0.006$ to $Z=0.013$ (Strobel \cite{strobel}), which, however, in terms
of masses, leads to an error of a few per cent at most. The CMDs, with
isochrones overplotted, are displayed in Figs.~\ref{tychocmds},
\ref{pleiadescmd}, and \ref{stock2cmds}. The parameters which are listed in
Table \ref{tychoisofit} were necessary for the transformation of the $V_T$
magnitudes to the initial stellar masses. From $(\alpha,\delta)$ plots of the
members we determined lower limits of the cluster diameters: We selected
diameters outside of which almost no more members are present, assuming that
these outliers are field stars with the same proper motion as the
corresponding cluster. These results are added to Table \ref{tychoisofit}. The
``true'' sizes might be larger due to mass segregation and the limited
magnitude range of our study.

\begin{table}
\caption[]{\label{tychoisofit} Distance modulus, reddening, and age
  estimations as determined during the isochrone fitting process. The
  determination of these parameters was necessary for the application of
  the mass-luminosity relation and therefore for the determination of the
  IMF. For a discussion of the differential reddening of Stock\,2, see
  Sect.~\ref{stock2}. Also lower limits $\underline{d}$ for the cluster
  diameters as derived from positional plots of the members are listed. The
  clusters marked with a star appear in vL99a}
\begin{tabular}{lccrc}
\hline
cluster & $(m-M)_0$ & $E_{B-V}$ & \multicolumn{1}{c}{$t$} & $\underline{d}$\\
 & [mag] & [mag] & \multicolumn{1}{c}{Myr} & [$\arcmin$]\\
\hline
Blanco\,1*    & $6.8$ & $0.03$ & $50$ & $105$ \\
Stock\,2      & $7.5$ & $\ldots$  & $100$ & $260$ \\
$\alpha$ Per* & $6.3$ & $0.09$ & $20$ & $255$ \\
Pleiades*     & $5.6$ & $0.05$ & $75$ & $300$ \\
NGC\,2451\,A* & $6.4$ & $0.00$ & $20$ & $140$ \\
IC\,2391*     & $5.8$ & $0.00$ & $20$ & $110$ \\
Praesepe*     & $6.0$ & $0.00$ & $650$ & $195$ \\
IC\,2602*     & $5.8$ & $0.03$ & $10$ & $185$ \\
NGC\,7092     & $7.6$ & $0.12$ & $70$ & $170$ \\
\hline
\end{tabular}
\end{table}

We found that the members of Stock\,2 are differentially reddened, whereas
the main sequences of the other objects either show a uniform reddening or are
too sparsely populated for a distinct proposition. The variations of the
reddening in the field of Stock\,2 will be discussed in detail in
Sect.~\ref{stock2}. 

\begin{figure}
\centerline{
\includegraphics[width=\hsize]{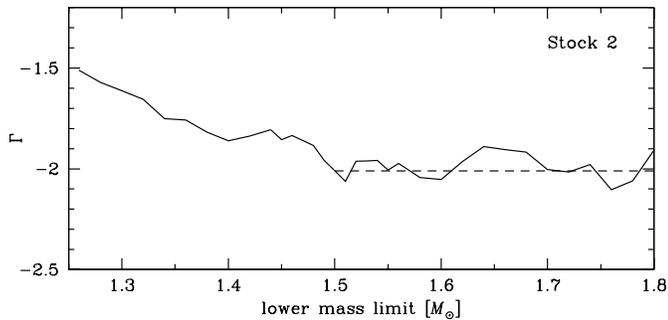}
}
\caption[]{\label{stock2gammas} IMF slope $\Gamma$ as derived for Stock\,2
  with our maximum likelihood technique depending on the adopted lower mass
  limit. The dashed line illustrates the final result of $\Gamma=-2.01$. We
  interpret this plot such that the data become incomplete around $m=1.5
  M_\odot$, i.e. $V_T=11.4 \mbox{~mag}$. See Sect.~\ref{tychodata} for a
  detailed explanation}
\end{figure}

We computed the masses from the $V_T$ magnitudes with the help of the
mass-luminosity relation provided together with the isochrones. We used
$6^{\rm th}$ order polynomials as an approximation. Attempts with polynomials
of lower orders failed since the low masses were not sufficiently well
reproduced. A comparison of our isochrones basing on the Bono et~al.\
(\cite{isoteramo}) models with the Geneva isochrones (Schaller et~al.\
\cite{schaller}) revealed that the difference between the masses as functions
of the $V$ magnitudes are less than $3 \%$. We derived the IMF slopes of
the clusters from the masses of the members with a maximum likelihood
method, i.e. we maximised the likelihood function
\begin{equation}
{\cal L}(\Gamma)= \prod_{i=1}^{n} \frac{m_i^\Gamma}{\int_{m_l}^{m_u}
  m^\Gamma \mbox{d} \log m}
\end{equation}
with the determined masses $m_i$ of $n$ stars and the limits of the normalising
integral being the lower and upper mass limits of the observed part of the
main sequence, $m_l$ and $m_u$. (For a discussion of the advantages of this
method, see Sanner et~al.\ \cite{capaper}.) A critical point of this work was
the completeness of the Tycho--2 data which would have contaminated the
results towards too high $\Gamma$ values, i.e. too shallow slopes. H{\o}g
et~al.\ (\cite{tycho2}) state in their Table 1 that the Tycho--2 data are 99
\% complete down to $V_T=11 \mbox{~mag}$ and still 90 \% down to $V_T=11.5
\mbox{~mag}$. However, it cannot be excluded that these figures vary for
different fields, depending on crowding or the first epoch material used to
build Tycho--2. We tried to get an idea about the completeness of the data by
the optical impression of the CMDs and by computing the IMF for different mass
ranges: We assumed that the IMF indeed has the shape of a single power
law within the range of the Tycho--2 photometry. If the lower mass limit is
below the completeness limit, the resulting IMF would be too shallow, since
stars at the low mass end of the function would be missed. When we vary the
lower mass limit up towards the completeness limit, $\Gamma$ would approach its
``correct'' value. Above this point, $\Gamma$ would remain on a stable
level. This phenomenon is illustrated with the Stock\,2 data in
Fig.~\ref{stock2gammas}. An alternate interpretation of this behaviour of the
slope as a function of the lower mass limit is that the shape of the IMF
deviates from a power law with constant slope, e.g. due to a gap in the main
sequence -- a phenomenon which is not uncommon for open star cluster CMDs
(see, e.g., the recent work of Rachford \& Canterna \cite{rachford} and
references therein). Therefore, the assumption that the IMF indeed follows a
``perfect'' power law may lead to underestimations of the completeness limit:
read the figures given in Table \ref{tychoergebnisse} as upper limits (in
terms of magnitudes) for the completeness limit in the fields of the
clusters. Both interpretations recommend us to stop the determination of the
IMF at the point where the slope becomes variable: We would either compute
the IMF in a mass interval in which the data are incomplete, or we would fit a
single power law to an IMF which does not follow such a function.

We found varying completeness limits by this investigation. For two of
the clusters, the star counts drop significantly at magnitudes as bright as
$V_T \approx 10 \mbox{~mag}$ (see Table \ref{tychoergebnisse}).

Most of the clusters are sparsely populated in the mass range under
consideration. This made the choice of the correct completeness limit a
challenging task. Fig.~\ref{tychoimfs} shows an overview of the IMFs of six of
the clusters; the Pleiades IMF is sketched separately  in
Fig.~\ref{pleiadesimfs}. The results are discussed in the following section
and summarised in Table \ref{tychoergebnisse}.

An alternative way of determining the completeness limit would be the
comparison of the Tycho--2 data with a second catalogue with fainter limiting
magnitude, like the GSC 1.2 (R\"oser et~al.\ \cite{gsc}) or USNO A2.0 (Monet
et~al.\ \cite{usno}): As long as no stars of the other catalogue are missing
in Tycho--2, we can assume Tycho--2 to be complete. However, for this
procedure it is required that the brightest stars missing in Tycho--2 are
present in the second catalogue. This in general cannot be expected, since
stars might be missing in both catalogues due to the presence of close
neighbours, crowding, or similar effects. With this method we would only be
able to obtain a limit at which Tycho--2 {\it definitely} is incomplete, which
is not enough for a thorough IMF study.

Adding the supplement 1 of the Tycho--2 catalogue (see H{\o}g et~al.\
\cite{tycho2} for details on the two supplements) did not affect the
completeness in general: Although we found from 7 (Blanco\,1) to 141 ($\alpha$
Per) stars in supplement 1, only less than 10 stars per cluster (except for
$\alpha$ Per, with 13 objects) could be found for which proper motions were
given. No more than four of them in each field also contained both $B_T$ and
$V_T$ photometry. For the stars in Tycho--2 and its supplement 1 which are
listed without proper motions, we also attempted to use the proper motions
given in other Hipparcos-based catalogues, namely ACT (Urban et~al.\
\cite{act}) and TRC (H{\o}g et~al.\ \cite{trc}). For the Pleiades, this
resulted in 52 additional stars. 21 of them can be considered as members
according to their proper motions, but only 7 objects could be added to the
IMF computation. The others either were located well below our completeness
limit (4 stars), or $B_T$ was not given (4 stars), or the data points were so
far away from the main sequence that we can assume that these data suffer from
bad photometry or the stars are non-members with the same proper motions as
the Pleiades (6 stars). An IMF computation with the 7 stars added led to a
difference in $\Gamma$ of $0.02$ compared with the IMF based on the Tycho--2
proper motions only. This difference is so much smaller than the errors that
we did not follow this procedure for the other clusters.

Comparing the proper motion centres of the clusters with the values
determined by Robichon et~al.\ (\cite{robichon}) using Hipparcos data, we
found that the results coincide very well in both $\mu_\alpha \cos \delta$ and
$\mu_\delta$ within $1 \mbox{~mas~yr}^{-1}$. This indicates that the
absolute proper motions of Tycho--2 correspond with the Hipparcos
catalogue and that at least in the fields of the open clusters studied there
are no systematic errors caused, e.g., by the first epoch material.

\section{Results for the individual clusters} \label{tychoindiv}

\subsection{Pleiades} \label{pleiades}

The first object under investigation is the Pleiades cluster. The Pleiades
show a proper motion centre which is located more than $35
\mbox{~mas~yr}^{-1}$ away from the average field proper motion. In addition,
previous IMF studies are available, so that we can externally check the
quality of our results.

In the case of the Pleiades, the standard deviation of the member proper
motions might not only reflect the inaccuracy of the Tycho--2 catalogue, but
also intrinsic properties of the Pleiades, like internal velocity
dispersion or perspective effects. However, since the members under
consideration are located in a comparably small area (see Table
\ref{tychoisofit}), these effects are of minor importance.

\begin{figure}
\centerline{
\includegraphics[width=\hsize]{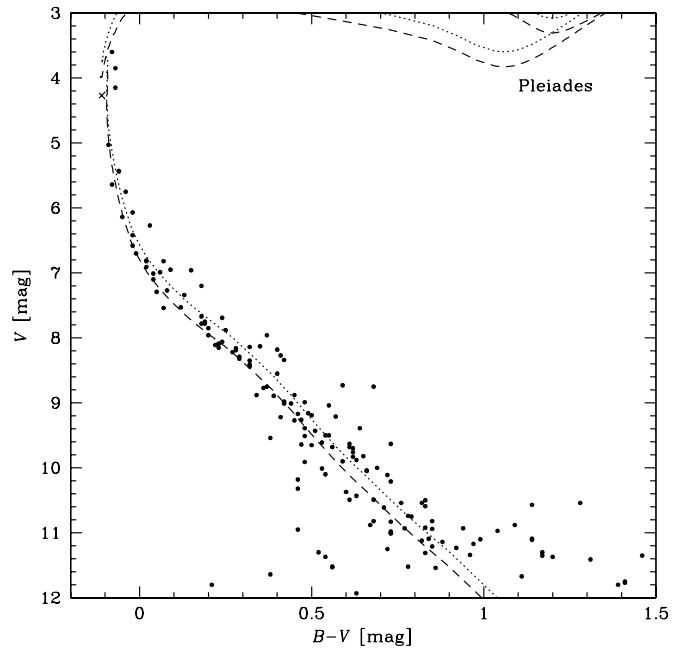}
}
\caption[]{\label{pleiadescmd} Colour magnitude diagram of the Pleiades
  members. We overplotted the Bono et~al.\ (\cite{isoteramo}) isochrones for
  $(m-M)_0=5.37 \mbox{~mag}$ (dotted line) and $(m-M)_0=5.6 \mbox{~mag}$
  (dashed line). Both isochrones are reddened by $E_{B-V}=0.05 \mbox{~mag}$. The star marked with a cross is listed as a blue straggler candidate
  in Ahumada \& Lapasset (\cite{bluestrag}). This object is discussed in the
  text (Sect.~\ref{pleiades})} 
\end{figure}

Fitting isochrones to the CMD of the members, we found that a distance
modulus of $(m-M)_0=5.37 \mbox{~mag}$ as given by vL99a is not the best
parameter in terms of quality of the fit: The isochrone passes right in the
middle of the main sequence, whereas it is our strategy to fit the isochrone
towards the blue side of the main sequence to account for binary stars or Be
stars, which broaden the ``true'' main sequence towards the red. It seems that
an isochrone with $(m-M)_0=5.6 \mbox{~mag}$ provides a better fit. For both
isochrones, a reddening of $E_{B-V}=0.05 \mbox{~mag}$ and an age of the
Pleiades of 75 Myr were assumed. The CMD with both isochrones overplotted is
shown in Fig.~\ref{pleiadescmd}. There has been a recent discussion about the
reliability of the Hipparcos parallaxes on which vL99a based his distance
determinations: Pinsonneault et~al.\ (\cite{pinsonneault}) as well as
Narayanan \& Gould (\cite{narayanan}) claim that small systematic errors in
the Hipparcos parallaxes occur, especially in the field of the Pleiades, so
that they derived a distance which corresponds to our value, while vL99a
excludes these errors as a consequence of extensive tests of the Hipparcos
data (see also van~Leeuwen \cite{vl99_2}).

\begin{figure}
\centerline{
\includegraphics[width=\hsize]{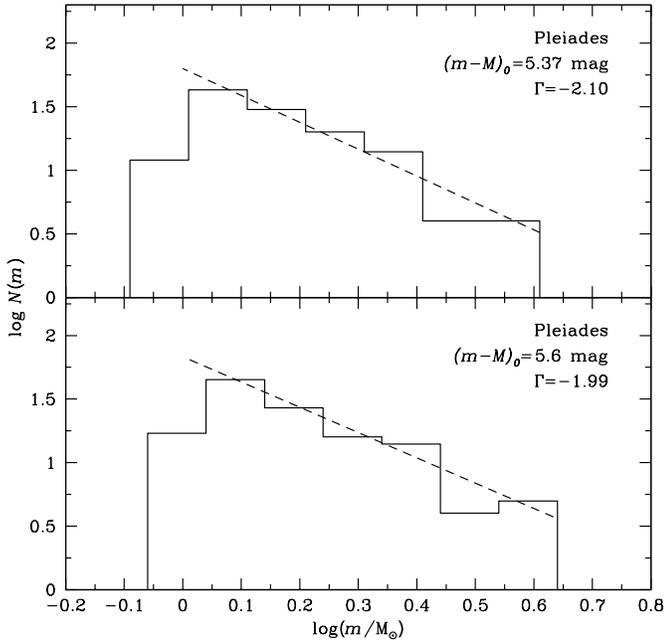}
}
\caption[]{\label{pleiadesimfs} Initial mass functions of the Pleiades
  cluster. The upper diagram is based on a distance modulus of $(m-M)_0=5.37
  \mbox{~mag}$, the lower one on $(m-M)_0=5.6 \mbox{~mag}$. The dashed lines
  symbolise the power laws with slopes of $\Gamma=-2.10$ and $\Gamma=-1.99$,
  respectively, the limits of the lines the mass interval under
  consideration. Note that, although we plot an underlying histogram for
  illustration, the functions were determined with a maximum likelihood
  method. The rightmost bins contain only a few stars, so that the bins
  in this range are not meaningful}
\end{figure}

\begin{figure}
\centerline{
\includegraphics[width=\hsize]{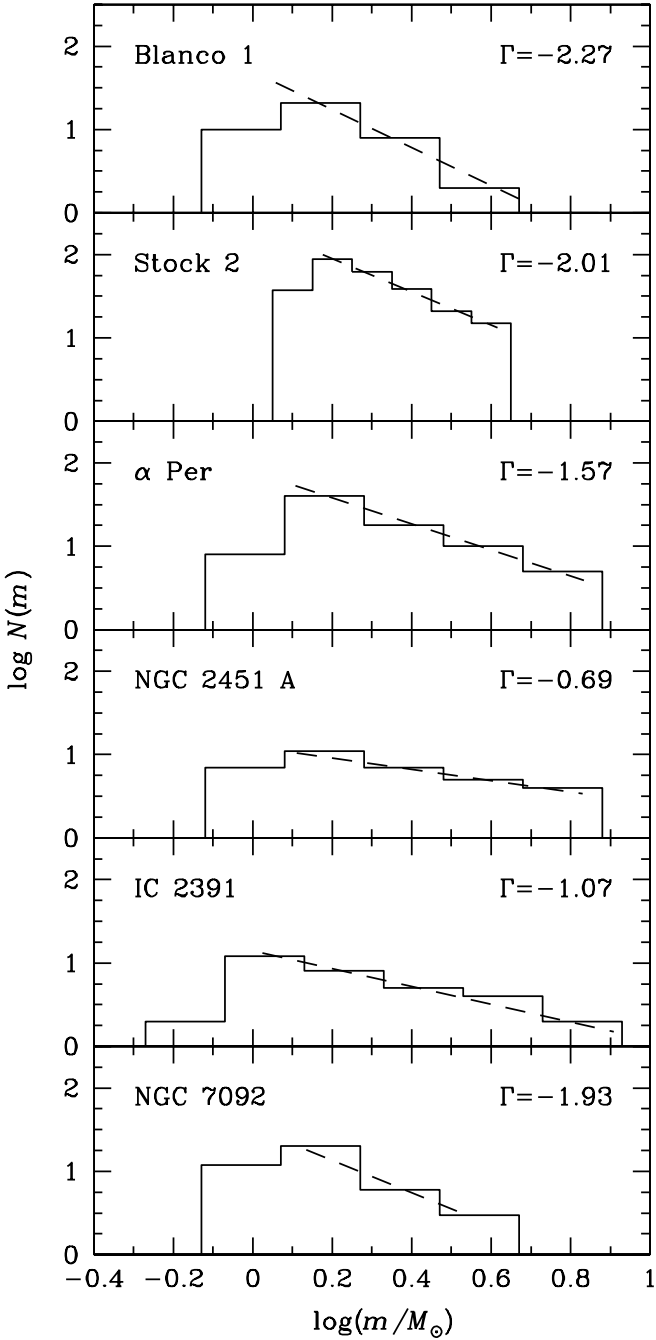}
}
\caption[]{\label{tychoimfs}Initial mass functions of six of the open star
  clusters using Tycho--2 data. The Pleiades IMF can be found in
  Fig.~\ref{pleiadesimfs}. For Praesepe and IC\,2602, no reliable IMF could be
  determined and their histograms are not shown. For more details about the
  plots, see the explanations in Fig.~\ref{pleiadesimfs}}
\end{figure}

We cannot resolve this problem with our results, therefore we computed the
IMF based on both parameter sets leading to results of $\Gamma= -1.99\pm 0.39$
and $\Gamma=-2.10 \pm 0.39$, with our and vL99a's distances, respectively. This
means that within the errors the IMF determination does not depend on the
uncertainties of the distance determination. The two mass functions are
sketched in Fig.~\ref{pleiadesimfs}. In Tables \ref{tychoisofit} and
\ref{tychoergebnisse} we retain our values.

In a study of the IMF of the Pleiades, almost in the same mass range as ours,
van~Leeuwen (\cite{vLpleiades}) arrived at a result of (transformed to our
notation) $\Gamma=-1.71 \pm 0.27$ based on the members known at that
time. Within the errors, this is in good agreement with our values.
Meusinger et~al.\ (\cite{meusinger}) studied the IMF down to approx. $0.4
M_\odot$ and stated that globally the Pleiades IMF cannot be represented by a
single power law. Therefore they did not provide a specific slope for $m \ga
M_\odot$, but they noted their IMF as being ``somewhat steeper'' than
van~Leeuwen's (\cite{vLpleiades}) result in the corresponding mass interval,
so that their findings seem to support our result very well.

A region with a radius of $7^\circ$ around the centre of the Pleiades contains
397 stars which are common to both the Hipparcos and Tycho--2
catalogues. This gave us a good opportunity to compare the proper motions of
the two catalogues. We found a good agreement with $\Delta \mu_\alpha \cos
\delta=-0.13 \pm 2.2 \mbox{~mas~yr}^{-1}$, $\Delta \mu_\delta=0.001 \pm 2.0
\mbox{~mas~yr}^{-1}$, so that the absolute reference systems of both
catalogues appear to be almost identical.

One of the three stars mentioned in the list of Ahumada \& Lapasset
(\cite{bluestrag}) as blue stragglers of the Pleiades can be found in the
Tycho--2 main catalogue; the other two are too bright, so that no proper
motions were determined for them. The object included in Tycho--2, HD\,23338,
has a proper motion of $\mu_\alpha \cos \delta= +20.1 \mbox{~mas~yr}^{-1}$,
$\mu_\delta= -44.8 \mbox{~mas~yr}^{-1}$, corresponding to a membership
probability of $P=0.98$, so that there is clear evidence that this object is a
member of the Pleiades. We marked this star in the CMD in
Fig.~\ref{pleiadescmd} with a cross. The star is located to the left of the
best fitting isochrone, however it is still well below the turn-off point of
the main sequence so that we do not see evidence for the blue straggler nature
of this object. For the other two objects, we can refer to the Hipparcos
proper motions instead: As stated before, the proper motions of Hipparcos and
Tycho--2 are almost identical, so that the pure Hipparcos figures can be taken
as an indication about membership without any transformation or other special
treatment, although we do not give explicit membership probabilities. In the
case of the Pleiades, the two stars not in Tycho--2 are HD\,23302 (HIP\,17499)
and HD\,23630 (HIP\,17702). Their proper motions are close to the centre of
proper motions of the Pleiades so that we can assume that these objects indeed
belong to the Pleiades. Note that since these two objects do not appear in the
Tycho--2 catalogue with complete sets of data, we did not include them in our
membership analysis and therefore they are not present in the CMD of the
Pleiades members (Fig.~\ref{pleiadescmd}).

Slettebak (\cite{slettebak}) provides data on Be and shell stars in open star
clusters. He reports four members of the Pleiades. The proper motions of
two of these stars, HD\,23302 (17\,Tau) and HD\,23630 ($\eta$ Tau) are not
listed in Tycho--2, but are in the Hipparcos catalogue. Both objects are likely
members of the Pleiades according to their Hipparcos proper motions. The two
remaining stars of the Slettebak (\cite{slettebak}) list, HD\,23480 (23\,Tau)
and HD\,23862 (28\,Tau) are cluster members with membership probabilities of
$P=0.97$, so that Tycho--2 supports the proposition that these objects are
members of the Pleiades, too. It is interesting that HD\,23302 appears in both
lists. However, it should be remarked that its category in Ahumada \& Lapasset
(\cite{bluestrag}) is 3, i.e. only weak evidence for a blue straggler nature,
whereas Slettebak's (\cite{slettebak}) discussion seems to be more sure.

\subsection{Stock\,2} \label{stock2}

Stock\,2 is by far the most populous cluster (down to the limiting magnitude
of Tycho--2) of our sample. There has been a controversial discussion about the
distance modulus and reddening of the cluster (see the overview in the
recently published work of Foster et~al.\ \cite{stock2ccd}). Here we just
summarise that distance moduli from $7.4 \mbox{~mag}$ (Piskunov
\cite{piskunov}) up to $8.36 \mbox{~mag}$ (Pandey et~al.\ \cite{pandey}) are
proposed; the reddenings found are of the order of $E_{B-V} \approx 0.3
\mbox{~mag}$.

\begin{figure*}
\centerline{
\includegraphics[width=\textwidth]{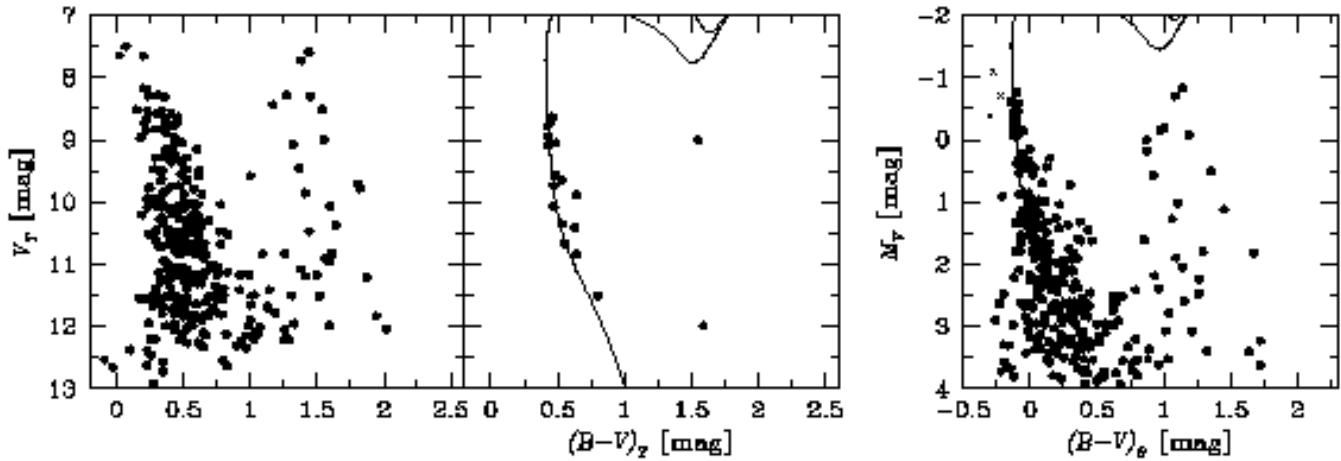}
}
\caption[]{\label{stock2cmds}Colour magnitude diagrams of the members of
  Stock\,2. The leftmost diagram shows all members with their apparent
  magnitudes.  Note the broad main sequence which inspired us to study the
  differential reddening of the cluster. The middle diagram shows the subfield
  of Stock\,2 which is accentuated in Fig.~\ref{stock2roet}. This field
  provided the evidence for the distance modulus of Stock\,2, determined
  as $(m-M)_0=7.5 \mbox{~mag}$. Overplotted is the isochrone which best fit
  these data, the reddening is $E_{B-V}=0.55 \mbox{~mag}$, the adopted
  age $t=100 \mbox{~Myr}$. The right panel is the de-reddened CMD
  of Stock\,2 with absolute magnitudes. We overplotted the $100 \mbox{~Myr}$
  isochrone. The red ($(B-V)_T \ga 1 \mbox{~mag}$) stars are most likely
  mis-identified field stars which by chance have the same proper motion as
  the cluster. The blue stars marked with crosses are either caused by the
  same factor or have a smaller reddening than the the region they are
  assigned to in our rough reddening map (Fig.~\ref{stock2roet})}
\end{figure*}

\begin{figure}
\centerline{
\includegraphics[width=\hsize]{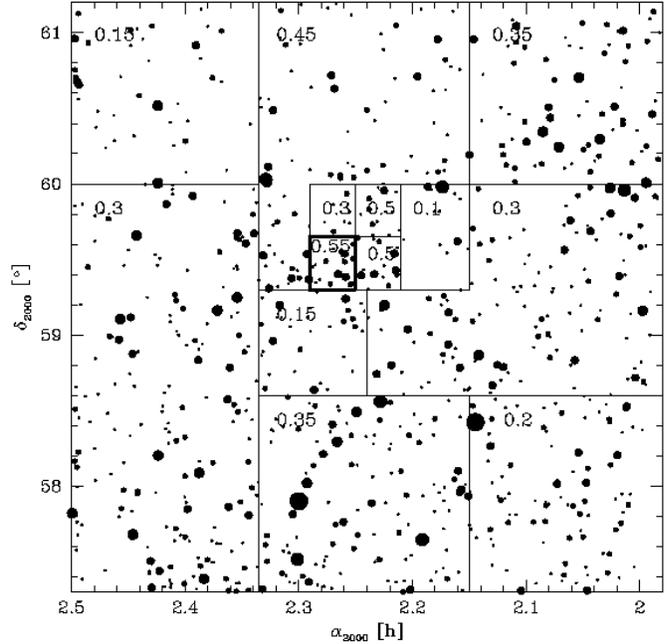}
}
\caption[]{\label{stock2roet} Reddening map of Stock\,2. The largest dots
  represent $V_T<6 \mbox{~mag}$ stars, the smallest ones are still brighter
  than $V_T=11 \mbox{~mag}$. North is up, east to the left. The fields could
  not be chosen arbitrarily small in size, as it had to be taken into account
  that the fields cover a sufficient amount of stars to see the
  relevant features of the main sequence. The CMD of the field marked with
  heavy borders and a reddening of $E_{B-V}=0.55 \mbox{~mag}$ is plotted in
  Fig.~\ref{stock2cmds}, middle diagram. This field turned out to be important
  for the distance determination of Stock\,2 (see Sect.~\ref{stock2})}
\end{figure}

Our CMD of the members (Fig.~\ref{stock2cmds}, left diagram) shows a broad
($\approx 0.3 \mbox{~mag}$) main sequence over the entire magnitude interval,
so that we decided to study the CMDs of smaller areas to find out whether
Stock\,2 is differentially reddened. We found that indeed for small portions
of the data the main sequences are much narrower, and we found different
reddening values: Large portions of the field of Stock\,2 are reddened by
$E_{B-V}=0.3 \mbox{~mag}$, however, we found areas with reddenings from
$E_{B-V}=0.1 \mbox{~mag}$ up to $0.55 \mbox{~mag}$. The results of this
investigation are presented in the reddening map of Stock\,2
(Fig.~\ref{stock2roet}). Since the single fields do not cover many objects, due
to the bright limiting magnitude of Tycho--2, a much more detailed reddening
map could be achieved with a deeper photometric study of Stock\,2.

While for most areas isochrones with a large range of distance moduli ($7.5
\mbox{~mag} \la (m-M)_0 \la 8.2 \mbox{~mag}$) led to good fits, one field
($2^{\rm h} 14^{\rm m} 24^{\rm s} \le \alpha_{2000} \le 2^{\rm h}
15^{\rm m} 0^{\rm s}$, $59^\circ 18 \arcmin \le \delta_{2000} \le
59^\circ 39 \arcmin$, marked in Fig.~\ref{stock2roet}) could not be
sufficiently well fitted after applying a distance modulus of more than
$(m-M)_0=7.5 \mbox{~mag}$. The CMD of this field together with the
corresponding isochrone is sketched in Fig.~\ref{stock2cmds}, middle
diagram. On the other hand, {\it all} the other fields do correspond to a
distance modulus of $(m-M)_0=7.5 \mbox{~mag}$. Therefore we adopt this
value to represent the distance modulus of Stock\,2. The right diagram of
Fig.~\ref{stock2cmds} shows the composition of the de-reddened CMDs of the
individual sub-fields on the basis of absolute magnitudes. It is evident that
the main sequence is much more pronounced than the original CMD (left
diagram in Fig.~\ref{stock2cmds}), at least down to $M_V=2 \mbox{~mag}$. With
a more detailed reddening map the CMD could still be improved.

The CMD shows three bright very blue ($(B-V)_T<-0.2 \mbox{~mag}$, marked in
the right diagram of Fig.~\ref{stock2cmds}) as well as numerous red ($(B-V)_T>1
\mbox{~mag}$) stars that do not fit to the selected isochrone. These objects
can be explained as randomly having the same proper motions as Stock\,2, but
in fact being field stars. The objects are uniformly distributed over the
field in an $(\alpha, \delta)$ plot; for example, two of them can be found in
the CMD of the selected region of Stock\,2 mentioned before
(Fig.~\ref{stock2cmds}, middle diagram). Another explanation for the blue
objects would be that they are less reddened than proposed for the field in
which they are located.

The IMF was computed from 262 main sequence stars within a field of view of
$4^\circ \times 4^\circ$, after some 70 too red and blue stars discussed
before had been removed. 204 objects are located above the completeness limit
of $V_T=11.0 \mbox{~mag}$, so that the slope of $\Gamma=-2.01 \pm 0.40$
finally is based on these stars.

We had mentioned in Sect.~\ref{tychodata} that the standard deviation of the
members of Stock\,2 was larger than for all the other clusters and higher than
expected from the general accuracy of the Tycho--2 catalogue. Stock\,2 is
located within the Vatican declination zone ($+55^\circ$ to $+64^\circ$) of
the Astrographic Catalogue. A remark of Eichhorn (\cite{eichhorn}, p. 297)
that the accuracy of the Vatican measurements are comparably low could explain
this substandard behaviour.

Five stars in the field of Stock\,2 are mentioned by Ahumada \& Lapasset
(\cite{bluestrag}) as blue stragglers belonging to the cluster. Three of
them -- their stars 6, 31 (HD\,13402), and 45 -- show membership
probabilities of $P=0$. The other two -- stars 110 (HD\,13735) and 138
(HD\,13909)-- have high probabilities of $P \approx 0.97$, however, after the
de-reddening process, they are located well in the main sequence of the
CMD. Therefore, we propose that all five objects are not blue straggler
members of Stock\,2.

\subsection{Praesepe (NGC\,2632, M\,44)}

Astrometric and photometric aspects of Praesepe have been well studied before
(e.g. Jones \& Cudworth \cite{praesepeeb}, Jones \& Stauffer
\cite{praesepeubv}), including IMF studies (Williams
et~al.\ \cite{praesepeimf}, Pinfield et~al.\ \cite{pinf}).

The Tycho--2 stars which were determined as members of Praesepe mostly are
located within a distance of $1.5^\circ$ of the centre of the cluster. The CMD
shows a scattered main sequence, so that a differential reddening seems
likely. Since the total number of detected member stars brighter than $V_T=10
\mbox{~mag}$ is only 79, a meaningful reddening map as determined for Stock\,2
(Sect.~\ref{stock2}) could not be obtained. The isochrone shifted according to
the vL99a results fits the main sequence quite well, however, the isochrone
lies slightly too red, so that we propose a reddening of $E_{B-V}=0.02 \mbox{~mag}$.

Due to the age of Praesepe, which we determined to be 650 Myr, only a small
range of main sequence stars could be covered. Therefore, it was not
possible to derive a trustworthy IMF from the Tycho--2 data.

Ahumada \& Lapasset's (\cite{bluestrag}) catalogue of blue stragglers contains
five stars in the region of Praesepe. Three of them, HD\,73618, HD\,73666
(40\,Cnc), and HD\,73819 have proper motions leading to high membership
probabilities ($P>0.9$). HD\,73711 shows an intermediate probability of
$P=0.79$, but it is still likely that this object belongs to Praesepe. The
final object, HD\,73120, is a clear non-member with $P=0$.

\subsection{$\alpha$ Per (Melotte\,20)}

Although numerous photometric and kinematic studies are available for the
$\alpha$ Per cluster (e.g. Mitchell \cite{alphaperubv}, Fresneau
\cite{fresneau}, Prosser \cite{prosser1,prosser2}), the IMF of this object has
not been investigated since the work of Tarrab (\cite{tarrab}).

\begin{figure}
\centerline{
\includegraphics[width=\hsize]{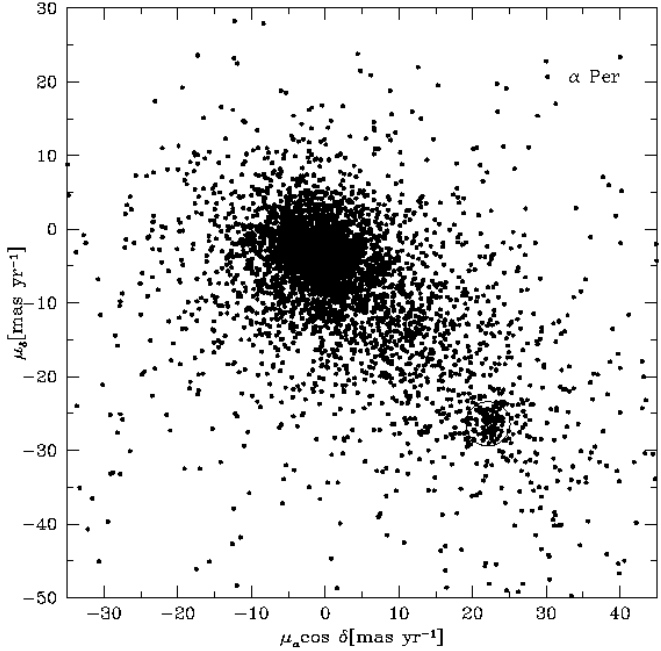}
}
\caption[]{\label{alphapervppd} Vector point plot diagram of the stars within
  $5^\circ$ of the centre of the $\alpha$ Per cluster. The cluster is
  clearly visible in the lower right part of the diagram. A circle is plotted
  around the region inside which the stars are considered members. This figure
  contains 4460 objects, 81 of which are determined to belong to the $\alpha$
  Per cluster}
\end{figure}

The vector point plot diagram of the $\alpha$ Per cluster is presented in
Fig.~\ref{alphapervppd}. The good separation between field and cluster proper
motions is clearly visible: Only very few field stars might erroneously be
identified as cluster members. The width of the distribution of the cluster
stars is $\sigma \approx 1.5 \mbox{~mas~yr}^{-1}$ which coincides with the
proposed standard deviation for individual objects included in the Tycho--2
catalogue. A circle indicates which stars were selected as the members of the
$\alpha$ Per cluster.

vL99a found values of $(m-M)_0=6.31 \mbox{~mag}$ and $E_{B-V}=0.09
\mbox{~mag}$ for the $\alpha$ Per cluster. We could confirm these values with
the Tycho--2 data. On the other hand, together with the proposed age of $50
\mbox{~Myr}$, the upper part of the main sequence is not well represented by
the fitted isochrone. We prefer a younger isochrone with $t=20 \mbox{~Myr}$,
which is plotted in Fig.~\ref{tychocmds}. A higher age could only be explained
by the brightest members detected being blue stragglers.

Three stars are listed as blue stragglers by Ahumada \& Lapasset
(\cite{bluestrag}). Two of them, HD\,22928 and HD\,24760, are too bright to be
included in the Tycho--2 catalogue with proper motions. One of them,
HD\,21428,  has a membership probability of $P=0.34$ according to its Tycho--2
proper motion, so that it can be assumed that this object is a field star. The
proper motions of the other two, HD\,22928 (HIP\,17358) and HD\,24760
(HIP\,18532), can again be taken from the Hipparcos catalogue: Both have a
proper motion which indicates they are not members of $\alpha$ Per, since for
HD\,22928 the deviation of $\mu_\delta$ from the cluster centre is approx. $20
\mbox{~mas~yr}^{-1}$, and for HD\,24760, $\mu_\alpha \cos \delta$ is too far
away from the value for the members.

In Slettebak (\cite{slettebak}), three objects are listed as Be or shell stars
in the region of the $\alpha$ Per cluster. HD\,21551 has a membership
probability of $P=0.39$ and HD\,22192 ($\psi$ Per) of $P=0.27$ which means
that both objects are likely non-members. HD\,25940 (48\,Per) is not included
in the Tycho--2 catalogue, but it is a Hipparcos star (HIP\,19343) with a
proper motion of $\mu_\alpha \cos \delta=20.19 \mbox{~mas~yr}^{-1}$,
$\mu_\delta=-33.26 \mbox{~mas~yr}^{-1}$. Comparing this with the proper motion
centre of the cluster (Table \ref{tychohaufen}), this means that this star,
too, shows a proper motion which does not coincide with the motion of the
cluster. Only the fourth star, HD\,21455, mentioned in the text of Slettebak
(\cite{slettebak}) as a non-member, has a membership probability of $P=0.76$
which makes it a likely cluster member from the proper motion point of view,
since the probability at which we distinguished members and non-members in the
case of $\alpha$ Per was $P=0.6$. Since Slettebak (\cite{slettebak})
mentions rotational velocities from Kraft (\cite {kraft}) and radial
velocities from Petrie \& Heard (\cite{petrie}) as further membership criteria,
it is still possible that this object in fact is not a member of the cluster.

\subsection{Blanco\,1} \label{blanco1}

\begin{figure}
\centerline{
\includegraphics[width=\hsize]{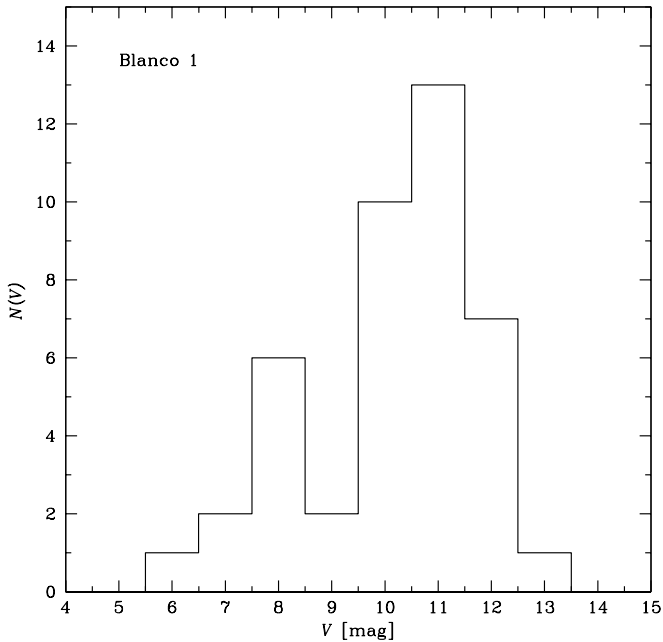}
}
\caption[]{\label{blanco1lkf} $V_T$ luminosity function of Blanco\,1. Note the
  dip around $V_T=9 \mbox{~mag}$}
\end{figure}

Blanco\,1 (also referred to as the $\zeta$ Scl cluster) turned out to be the
cluster with the second faintest completeness limit of our sample with
$V_T=11.4 \mbox{~mag}$, following the rough method presented earlier in
Sect.~\ref{tychodata}. Nevertheless, the IMF was computed on the basis of
34 stars only. The distance moduli of Blanco\,1 according to vL99a -- $6.82
\mbox{~mag}$ -- and Robichon et~al.\ (\cite{robichon}) -- $7.1 \mbox{~mag}$
-- differ by almost $0.3 \mbox{~mag}$. We found that an isochrone on the
basis of the vL99a value fits the CMD of the members slightly better, but
the Robichon et~al.\ (\cite{robichon}) distance modulus cannot be excluded,
either. Instead of a reddening of $E_{B-V}=0.09 \mbox{~mag}$, as proposed by
vL99a, we suggest a lower value of $E_{B-V}=0.03 \mbox{~mag}$. This value is
in good agreement with other studies of Blanco\,1, e.g. Westerlund
et.~al.\ (\cite{blanco1ubv}). The cluster age as given in vL99a could be well
reproduced. The luminosity function of Blanco\,1 (see Fig.~\ref{blanco1lkf})
shows a gap in the range from approx. $V_T=9 \mbox{~mag}$ to $V_T=9.7
\mbox{~mag}$ (corresponding to masses from $m= 1.6 M_\odot$ to $m=1.9
M_\odot$). When experimenting with the lower mass limit during the IMF
determination, this led to a temporary increase of $\Gamma$ from $-2.27$ to as
much as $-1.12$, which normalised after passing the gap. It must be remarked
that only 11 members have masses higher than $1.9 M_\odot$, so that the IMF can
most reliably be described with the help of the lower mass stars. The gap
seems not to be a consequence of incomplete Tycho--2 data, since it has been
observed before (Eggen \cite{eggen0}). In addition, these gaps are not
uncommon for open clusters (Rachford \& Canterna \cite{rachford}). As a
consequence, Blanco\,1 is an example of a star cluster for which a power law
does not well describe the shape of the entire IMF. This is also illustrated
by the comparably high error of the maximum likelihood fit. It would be
interesting to obtain a deeper photometric study of this object to find out
whether the IMF shows a ``normal'' power law shaped behaviour for lower mass
stars or if further gaps appear. An IMF computed on the basis of the
vL99a distance modulus leads to a slope of $\Gamma=-2.27 \pm 0.70$.

The star $\zeta$ Scl itself is, according to Ahumada \& Lapasset
(\cite{bluestrag}), a blue straggler cluster member. For this object we find a
membership probability of $P=0$ as a consequence of its proper motion of
$\mu_\alpha \cos \delta=-11.2 \mbox{~mas~yr}^{-1}$, $\mu_\delta=+5.8 \mbox{~mas~yr}^{-1}$.

\subsection{NGC\,2451\,A}

NGC\,2451 has been the subject of a discussion regarding whether it really is
an open star cluster or not (Eggen \cite{eggen1,eggen2}, R\"oser \& Bastian
\cite{roeser2451}). Recently, Carrier et~al.\ (\cite{carrier}) found
evidence that this object actually consists of two clusters in the same line
of sight, as proposed earlier by R\"oser \& Bastian
(\cite{roeser2451}). Baumgardt (\cite{baumgardt}) was only able to confirm
existence of one of the two clusters on the basis of the Hipparcos and ACT
catalogues. Carrier et~al.\ (\cite{carrier}) named the two objects
NGC\,2451\,A and B. We will stick to this nomenclature in the
following. Cluster A, which is the target of our study, is located at a
distance of approx. 190 pc, whereas component B is 360 pc ($(m-M)_0=7.8
\mbox{~mag}$) away from the Sun. NGC\,2451\,B will not be studied here,
as its proper motion is too close to the centre of the field proper motion.

For NGC\,2451\,A, we found that adopting the vL99a reddening of $E_{B-V}=0.04
\mbox{~mag}$ leads to an isochrone which is too far redwards to properly
represent the main sequence. vL99a does not distinguish between components A
and B. However, the values given for the proper motions (his Table 1) indicate
that he -- as well as Robichon et~al.\ (\cite{robichon}) -- really does
investigate component A. Carrier et~al.\ (\cite{carrier}) found the reddening
to be $E_{B-V}=0.01 \mbox{~mag}$, which better corresponds to our isochrone
fit. Due to the lack of evolved stars, the age determination is a critical
task: A $36 \mbox{~Myr}$ isochrone fits the data as well as a 10 Myr
isochrone. Carrier et~al.\ (\cite{carrier}) seem to overestimate the age of
NGC\,2451\,A with their value of 50 Myr.

With $\Gamma=-0.69$, NGC\,2451\,A shows the shallowest IMF of our
sample. Since the usable portion of the main sequence consists of 27 stars
only, this result is quite uncertain. Still, we can conclude that the IMF of
this object is not as steep as Scalo's (\cite{scalo2}) average slope of
$\Gamma=-1.7$ would suggest.

Ahumada \& Lapasset (\cite{bluestrag}) mention two stars in the field of
NGC\,2451 as blue stragglers. One of them, HD\,61831, is a likely
cluster member with a probability of $P=0.9$. However, since it is located
slightly right of the fitted isochrone in the CMD, its nature as a blue
straggler is doubtful. On the other hand, the second object, HD\,63465, has a
proper motion of $\mu_\alpha \cos \delta=-11.2 \mbox{~mas~yr}^{-1}$,
$\mu_\delta=+5.8 \mbox{~mas~yr}^{-1}$ which leads to a membership probability
of $P=0$. Following Carrier et~al.\ (\cite{carrier}) who indirectly give
evidence for the proper motion of cluster B with the statement that
HR\,2968 ($\mu_\alpha \cos \delta=-1.4 \mbox{~mas~yr}^{-1}$, $\mu_\delta=-3.9
\mbox{~mas~yr}^{-1}$) belongs to NGC\,2451 B, HD\,63465 does not seem to be a
member of either of the clusters.

\subsection{IC\,2391}

As for NGC\,2451\,A, IC\,2391 seems to be less reddened than proposed by
vL99a. Our result that IC\,2391 is unreddened can also be found, e.g.,
in Perry \& Hill (\cite{ic2391ubv}). The main sequence of IC\,2391 (see
Fig.~\ref{tychocmds}) is not very evenly populated: Several gaps of $\Delta
V_T \approx 0.7 \mbox{~mag}$ can be found around $V_T=6 \mbox{~mag}$, $V_T=7
\mbox{~mag}$, and $V_T=8.25 \mbox{~mag}$. Below $V_T=11 \mbox{~mag}$ only
very few stars with the proper motion of the cluster are present in the CMD,
and they are widely spread in colour. It seems that the data are complete down
to $V_T \approx 11 \mbox{~mag}$. IC\,2391 in general is sparsely populated, so
that we had to compute its IMF on the basis of only 29 stars. The result,
$\Gamma=-1.07 \pm 0.53$, again is at the shallow end of the IMF slope
interval.

\subsection{IC\,2602}

\begin{figure}
\centerline{
\includegraphics[width=\hsize]{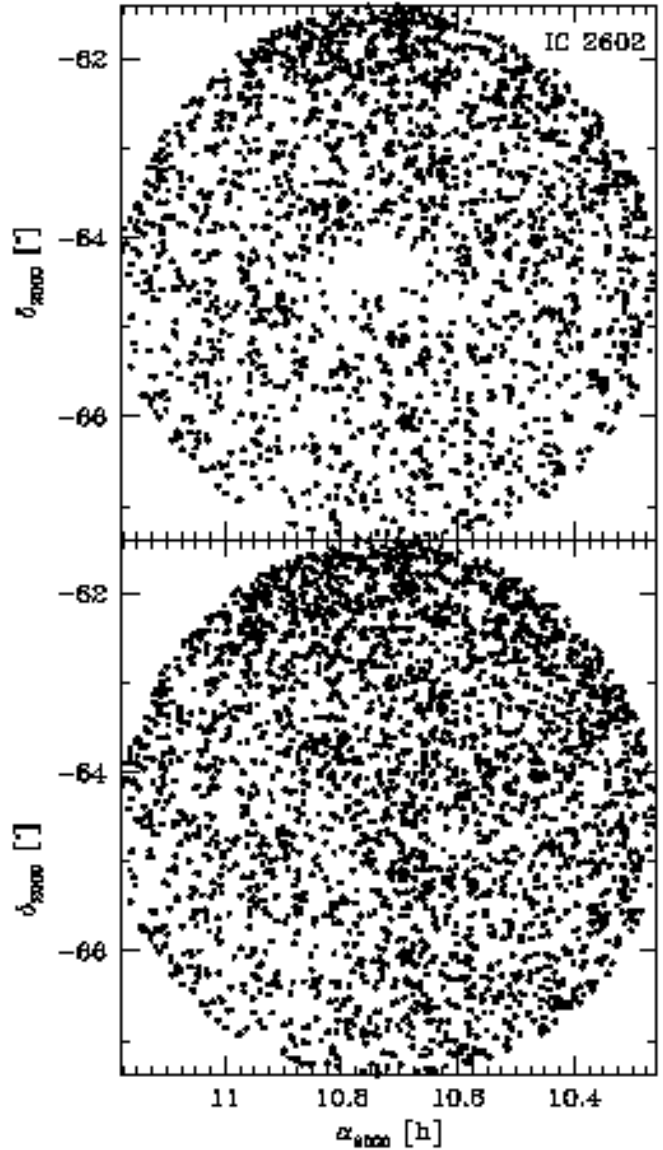}
}
\hfill
\caption[]{\label{ic2602dens} Plots of the stars in the region of IC\,2602 in
  the magnitude range $8 \mbox{~mag} \le V_T \le 11.5 \mbox{~mag}$ of the
  Tycho--2 stars (upper diagram) and GSC 1.2 (lower diagram). The radius of
  the field of view is $3^\circ$. Note the lack of stars in this magnitude
  range in the Tycho--2 data}
\end{figure}

IC\,2602 shows a clear and narrow main sequence down to $V_T=8
\mbox{~mag}$. Below this point, almost no stars are found, and the few detected members
are widely scattered in the CMD. In a positional plot of all stars with $8
\mbox{~mag} \le V_T \le 11.5 \mbox{~mag}$ of the Tycho--2 data (see
Fig.~\ref{ic2602dens}) one finds that in a $1^\circ$ wide
area around the centre of IC\,2602, almost no stars are found. An equivalent
plot of, e.g., the GSC, Version 1.2, (R\"oser et~al.\ \cite{gsc}) does not show
this decrease in stellar density. This might mean that for some reason the
stars fainter than $V_T=8 \mbox{~mag}$ in this area -- and hence members of
the cluster -- are not included in the Tycho--2 catalogue. The total stellar
density, as one can see in images of this region, is low enough so that neither
crowding nor neighbouring bright stars might have disturbed the measurements
of either the Tycho star mapper or the first epoch data. We also checked the
stars in Tycho--2 that are listed without proper motions and the two
supplements of Tycho--2. Supplement 1 contains 31 stars, but only for one of
them is a proper motion given. There is an above-average number of these stars
in the region of IC\,2602, but when we added all these objects, the total
stellar numbers are still too low. The original Tycho catalogue does not show
this lack of stars.

$V_T=8 \mbox{~mag}$ corresponds to a mass of $m=1.9 M_\odot$, and the
number of members which could be used for the IMF computation was no more
than ten, too small a number to compute the mass function of IC\,2602.

Ahumada \& Lapasset (\cite{bluestrag}) list a ``well known blue straggler''
for IC\,2602: $\vartheta$ Car. It indeed has a Hipparcos proper motion close
to the centre of the cluster proper motion.

\subsection{NGC\,7092 (M\,39)} \label{n7092sect}

For the sparse open cluster NGC\,7092, reddenings of a few hundredths of a
magnitude can be found in the literature (e.g., Mohan \& Sagar
\cite{n7092ubv}, Manteiga et~al.\ \cite{manteiga}, Nicolet \cite{nicolet}),
however, we found that no stars in the entire field studied (members {\it and}
non-members) justify an isochrone reddened by less than $E_{B-V}=0.1
\mbox{~mag}$. Concerning the distance modulus, our result of $(m-M)_0=7.6
\mbox{~mag}$ is in good agreement with other studies. The determination of the
IMF slope of NGC\,7092 led to a high error of $\Delta \Gamma >1$. This cannot
only be explained with the low number of 25 members above the estimated
completeness level, which, with $V_T=9.9 \mbox{~mag}$, is the brightest of the entire study besides IC\,2602. Another point is the shape of the IMF
which cannot be well represented by a power law: The CMD of NGC\,7092 shows a
magnitude region with (almost) no stars, in this case for $7.7 \mbox{~mag}
\la V_T \la 8.6 \mbox{~mag}$. In terms of masses, this interval corresponds
to $1.8 M_\odot \la m \la 2.4 M_\odot$. This again underlines the advantages
of a maximum likelihood estimation compared with a least square fit to an IMF
histogram, which would have led to an error of $\Delta \Gamma= \pm 0.14$ only,
so that the unusual shape of the IMF would not have been striking.

\section{Summary and discussion} \label{tychoconcl}

\begin{table}
\caption[]{\label{tychoergebnisse}IMF slopes $\Gamma$ of the star clusters
  under consideration together with the formal errors derived during the
  maximum likelihood analysis. Also the mass and $V_T$ magnitude ranges which
  could be taken into account are listed. The upper limit of the $V_T$
  interval can be taken for an estimation of the completeness limit of the
  Tycho--2 data in the regions of the clusters. For this point, see the
  remarks given in Sect. \ref{tychodata}. The numbers of stars given are
  valid for the members within these intervals}
\begin{tabular}{lrrcc}
\hline
cluster & \# & \multicolumn{1}{c}{$\Gamma$} & mass range & $V_T$ range\\
& stars & &[$M_\odot$] & [mag]\\
\hline
Blanco\,1 & $34$ & $-2.27 \pm 0.70$ & [$1.1; 4.8$] & [$6.1; 11.4$] \\
Stock\,2 & $204$ & $-2.01 \pm 0.40$ & [$1.5; 4.1$] & [$7.6; 11.0$] \\
$\alpha$ Per & $70$ & $-1.57 \pm 0.44$ & [$1.1; 6.8$] & [$5.0; 10.5$] \\
Pleiades & $127$ & $-1.99 \pm 0.39$ & [$1.0; 4.1$] & [$5.0; 10.9$] \\
NGC\,2451\,A & $27$ & $-0.69 \pm 0.63$ & [$1.3; 6.8$] & [$4.8; 10.0$] \\
IC\,2391 & $29$ & $-1.07 \pm 0.53$ & [$1.1; 8.1$] & [$3.5; 10.7$] \\
NGC\,7092 & $25$ & $-1.93 \pm 1.24$ & [$1.4; 3.4$] & [$6.5; 9.9$] \\
\hline
\end{tabular}
\end{table}

We used the information of the Tycho--2 catalogue for the determination of
the IMF of nearby open star clusters in the $\ga 1 M_\odot$ mass range and to
demonstrate the capabilities of Tycho--2.

The IMF of seven open star clusters could be represented by power laws with
exponents ranging from $\Gamma=-0.69$ to $\Gamma=-2.27$. Within the errors,
this is in good agreement with the range of IMF slopes as given, e.g., by
Scalo (\cite{scalo2}). An average slope from our results could be computed as
$\Gamma=-1.65$. The standard deviation of the sample of slopes equals $\Delta
\Gamma=0.6$, whereas the standard deviation as derived from the individual
errors of the slopes is only $\Delta \Gamma=0.26$.

Two of the clusters, Blanco\,1 and NGC\,7092, show significant gaps over
almost $1 \mbox{~mag}$ in luminosity or a few tenths of a Solar mass,
influencing the slope determined and their errors. For these clusters, further
studies with fainter limiting magnitudes will be needed to find a more precise
IMF, since not only would the mass range be larger, but also the numbers of
members to be detected.

With a Monte Carlo simulation, we tested whether our assumption that the IMF of
the clusters can be represented with a (single) power law within the mass
range covered by our study: We repeatedly simulated the mass functions of a
cluster with the same number of members as found from Tycho--2 and with the
slope derived from the data. From these samples, we determined the slopes with
the same method as for the actual objects. A probable difference between the
errors as found from the maximum likelihood method and the standard deviation
of the $\Gamma$ values found from the simulation might be an indication
that the IMF of that object does not follow a power law. All clusters
show an error smaller by $0.15$ in the simulated clusters. This could not only
be explained by the fact that the ``true'' IMF of the clusters does not follow
a power law, but also by the slight contamination of the data with field
stars (See the remarks in Sect. \ref{tychodata}). The influence of these
objects at arbitrary distances cannot be well simulated in the Monte Carlo
experiments. The only exception is NGC\,7092, with a difference of $0.4$. The
gaps in the main sequence mentioned in Sect. \ref{n7092sect} and the low
number of total stars might be the cause for this behaviour. However, we do
not find evidence in any of the clusters that the choice of a power law as
the representative of the IMF was clearly incorrect.

Comparing our results with the study of Tarrab (\cite{tarrab}), we find good
agreement for four of the five clusters which are discussed in both
publications. Only the slope of NGC\,7092, for which Tarrab found -- in our
notation -- $\Gamma=-0.72$, dramatically differs from our value of
$\Gamma=-1.93$.

Tycho--2 allows us to discuss the membership of individual objects
mentioned previously in the literature. We inspected the catalogue of blue
stragglers of Ahumada \& Lapasset (\cite{bluestrag}). Many of their objects
belong to their ``class 3'', which means that there is only weak evidence (like
the location in a CMD) for stars being blue stragglers {\it and} members of the
clusters. For these objects, if they are bright enough, proper motions are
now available, so that the list can be improved by cleaning it of field
stars. The second paper under consideration was Slettebak's (\cite{slettebak})
list of Be and shell stars in open clusters. While we could confirm the list
for the Pleiades, the nature of the stars mentioned in the context of the
$\alpha$ Per cluster has to be revised.

Our analysis showed that the completeness given for the Tycho--2 catalogue by
H{\o}g et~al.\ (\cite{tycho2}) is too optimistic, at least in several fields of
the clusters studied: A field with a diameter of approx. $1^\circ$ around the
centre of IC\,2602 contains just a few stars with $V_T>8 \mbox{~mag}$, in
contrast to other star catalogues. It should be remarked that this result is
not in contradiction with the statement about the global completeness of
Tycho--2 in H{\o}g et~al.\ (\cite{tycho2}). However, for studies in small
excerpts of the Tycho--2 data, one has to be aware of these regional variations
in the completeness. On the other hand, the quality of the Tycho--2 proper
motions appears to be as reliable as the Hipparcos data.

Regarding upcoming missions of astrometry satellites -- FAME (Horner
et~al.\ \cite{fame}), DIVA (R\"oser et~al.\ \cite{diva}), or
GAIA (Lindegren \& Perryman \cite{gaia}) -- studies based on uniform all-sky
data samples like the IMF investigations demonstrated in this work will play a
more important role in the future: The limiting magnitudes of these
instruments will be fainter, proper motions will be available for all detected
objects (in contrast to, e.g., the Hipparcos catalogue, the proper motions of
which were determined for the stars of an input catalogue only) and will
become independent of ground based observations, so that their data quality
will be higher. This means that -- concerning the determination of mass
functions -- the IMF of nearby clusters will be covered down to lower masses.
Due to the fainter limiting magnitude and the higher accuracy of the
proper motions (and hence a more precise membership determination) a larger
number of open star clusters can then be studied.

\begin{acknowledgements}
The authors acknowledge Santi Cassisi for providing the isochrones necessary
for our studies. For valuable discussions, we thank Andrea Dieball (especially
for the hints concerning the differential reddening and the study of the
stellar density) and Ram Sagar. Thanks to Klaas~S.\ de~Boer and Andrea Dieball
for carefully reading the manuscript of this publication. This research has
made use of NASA's Astrophysics Data System Bibliographic Services, the CDS
data archive in Strasbourg, France, and J.-C. Mermilliod's WEBDA database of
open star clusters.
\end{acknowledgements}

\end{document}